\documentclass[lettersize,journal]{IEEEtran}
\usepackage{amsmath,amsfonts}
\usepackage{algorithmic}
\usepackage{algorithm}
\usepackage{array}
\usepackage[caption=false,font=normalsize,labelfont=sf,textfont=sf]{subfig}
\usepackage{textcomp}
\usepackage{stfloats}
\usepackage{url}
\usepackage{verbatim}
\usepackage{graphicx}
\usepackage{cite}
\usepackage{listings}
\usepackage{xcolor}
\definecolor{codegreen}{rgb}{0,0.6,0}
\definecolor{codegray}{rgb}{0.5,0.5,0.5}
\definecolor{codepurple}{rgb}{0.58,0,0.82}
\definecolor{backcolour}{rgb}{0.95,0.95,0.92}

\newcommand\notype[1]{\unskip}

\lstdefinestyle{mystyle}{
    backgroundcolor=\color{backcolour},   
    commentstyle=\color{codegreen},
    keywordstyle=\color{magenta},
    numberstyle=\tiny\color{codegray},
    stringstyle=\color{codepurple},
    basicstyle=\ttfamily\footnotesize,
    breakatwhitespace=false,         
    breaklines=true,                 
    captionpos=b,                    
    keepspaces=true,                 
    numbers=left,                    
    numbersep=5pt,                  
    showspaces=false,                
    showstringspaces=false,
    showtabs=false,                  
    tabsize=2
}

\begin{document}

\title{Fostering the integration of European Open Data into Data Spaces through High-Quality Metadata}

\author{Javier Conde, Alejandro Pozo, Andr\'es Munoz-Arcentales, Johnny Choque and \'Alvaro Alonso
\thanks{J. Conde, A. Pozo, A. Munoz-Arcentales and \'A. Alonso are with the Departamento de Ingeniería de Sistemas Telemáticos, Escuela Técnica Superior de Ingenieros de Telecomunicación, Universidad Politécnica de Madrid. Email: javier.conde.diaz@upm.es}
\thanks{J. Choque is with the Departamento de Ingeniería de Comunicaciones, Escuela Técnica Superior de Ingenieros Industriales y de Telecomunicación, Universidad de Cantabria.}
}

\markboth{This work has been submitted to the IEEE for possible publication. Copyright may be transferred without notice.}
{Shell \MakeLowercase{\textit{et al.}}}



\maketitle

\begin{abstract}
The term Data Space, understood as the secure exchange of data in distributed systems, ensuring openness, transparency, decentralization, sovereignty, and interoperability of information, has gained importance during the last years. However, Data Spaces are in an initial phase of definition, and new research is necessary to address their requirements. The Open Data ecosystem can be understood as one of the precursors of Data Spaces as it provides mechanisms to ensure the interoperability of information through resource discovery, information exchange, and aggregation via metadata. However, Data Spaces require more advanced capabilities including the automatic and scalable generation and publication of high-quality metadata. In this work, we present a set of software tools that facilitate the automatic generation and publication of metadata, the modeling of datasets through standards, and the assessment of the quality of the generated metadata. We validate all these tools through the YODA Open Data Portal showing how they can be connected to integrate Open Data into Data Spaces.
\end{abstract}


\section{Introduction}
\IEEEPARstart{T}{he} current society is data-driven; in fact, terms such as data economy, digital transformation, data marketplace, have gained popularity in recent years~\cite{data_economy}. In the EU27 the data economy had a value of 325 billion euros, and most countries have initiatives or programs focused on data, such as the European Data Strategy led by the European Commission~\cite{european_strategy_for_data}, the Federal Data Strategy~\cite{federal_data_strategy}, the Office of Personnel Management Data Strategy~\cite{us_strategy_data} from the US, or the 14th Five-Year Plan for National Informatization from China~\cite{china_strategy_data}.  

The fundamental principles of the data economy are built on the value of data, interoperability, accessibility, and data sovereignty~\cite{Kirstein2022}. All these elements converge in the concept of Data Spaces, which represents an open and federated infrastructure for the secure exchange of data following common rules, standards, and policies~\cite{Ahle2022}. Data Spaces are a developing European initiative and their precursor is the Open Data (OD) ecosystem~\cite{reiberg2022data}. OD provide the facilities for data exchange, harvesting, and accessibility~\cite{reiberg2022data}. In its origin, OD emerged as a consequence of political corruption, however, they have resulted in a huge impact in the social and economic sectors. In 2018 OD accounted for 52 billion euros in European countries. By 2030, this quantity is estimated to increase to 194 billion euros~\cite{prediction_od}. 

OD have been used in numerous fields highlighting OD-based smart city applications~\cite{big_data_applications, future_trends_and_current_state}. They have also been used to develop solutions in many other fields such as digital twins~\cite{collaboration_of_dts}, cultural heritage~\cite{practices_lod_archeology}, agriculture~\cite{computational_science}, medicine~\cite{rutherford2022automated}, etc. Even OD has been used in artificial intelligence by providing datasets to train machine learning models~\cite{9963753, mahajan2023predicting}.

Despite their adoption, the OD ecosystem still faces barriers and challenges that need to be addressed for its adoption in Data Spaces. Specifically, ensuring that published data meet sufficient quality standards and adhere to FAIR principles (findability, accessibility, interoperability, and reusability), data sources require metadata to enable their discovery and these metadata have to follow standards and possess sufficient quality to facilitate its exchange.

In this study, we explore the barriers of OD and propose solutions to facilitate their integration into Data Spaces. To achieve this, we have developed a series of software components that enable the automation of data publication in Open Data Portals and the automation of generating metadata compatible with the Data Catalog Vocabulary (DCAT)~\cite{dcat_2} standard while ensuring their quality. The proposed tools have been validated in real scenarios through the Your Open DAta (YODA) Portal, which has been integrated into the European Data Portal resulting in the best Open Data Portal in Europe in terms of metadata quality to date (November 2023).

The document is structured as follows: in the following section, we analyze the state of the art of the OD ecosystem and its relationship with Data Spaces. In Section~\ref{ch:od_europe}, we examine the current state and barriers of OD in Europe and their role in the Data Space ecosystem. In Section~\ref{ch:automatic_publication}, we present our proposal for automatic publication and generation of high-quality metadata. Then, we validate the proposed solution through more than 200 datasets published on the YODA portal and aggregated on data.europa.eu. Finally, we present the conclusions and future work of our research.

\section{Related Work}
\label{ch:related}

\subsection{Origin of OD and Data Spaces}
In the early 2000s, due to political corruption, governments showed interest in publishing data accessible to all citizens. \IEEEpubidadjcol
The aim is to ensure the transparency of information~\cite{transparency_by_design_what_is, a_systematic_review}. These initiatives led to the emergence of so-called Open Data, i.e. data provided without restrictions on use, accessible to both people and machines.

In parallel, Tim Berners Lee defined the Semantic Web as an extension of the Web of Documents~\cite{the_semantic_web}. The Web of Documents is based on the access to information through HTML documents in which the relationship between documents is done through hyperlinks. The Semantic Web is an evolution of the Web of Documents based on a network of linked data (LD) that can be processed and self-discoverable by machines~\cite{linked_data_the_story}. Furthermore, if these data are offered publicly and through non-proprietary formats, they are called Linked Open Data (LOD). LOD promote access to information and increase the value of the data~\cite{linked_data_design}.

The European Union Directive 2003/98/EC~\cite{directive_2003} was issued in 2003. This directive mentions the value of public sector documents and the need to share them for their exploitation. Moreover, it points out the need to establish a common framework in the national rules that favor the search and reuse of documents. In 2013, Directive 2013/37/EU updated the 2003 version, which is outdated due to the rapid technological evolution and the rise of the data economy~\cite{directive_2013}. This directive talks about the need to offer documents in machine-readable formats, i.e. documents containing structured data that are processable by computer programs. It also mentions the importance of providing metadata that facilitates the search and identification of documents. A new revision of the directive was published in 2019 through Directive (EU) 2019/1024~\cite{directive_2019}. This directive mentions the value of dynamic data and the need to offer OD consumption mechanisms through APIs to provide the data in real-time.

The concept of OD emerged from the need for governments to demonstrate transparency. Although it has experienced significant growth over the years, most data providers remain public entities. The concept of Data Spaces emerged in 2005, in parallel to the growth of OD~\cite{10.1145/1107499.1107502}. Initially, Data Spaces were understood as a distributed ecosystem for the semantic data integration of different data sources through common vocabularies. The concept has evolved over the years and is nowadays understood as systems that enable the secure exchange of data in distributed systems, ensuring openness, transparency, decentralization, sovereignty, and interoperability of information~\cite{Ahle2022}.

The term gained importance in 2015 when the International Data Space initiative appeared with the aim of designing a distributed software architecture that guarantees data sovereignty~\cite{Otto2022_the_evolution}. In 2020, the EU established as one of its objectives the definition and development of data spaces through the European strategy for data~\cite{european_strategy_for_data}. Currently, the EU is in the phase of defining and implementing the basic components of Data Spaces. In 2021 the Data Spaces Business Alliance (DSBA)\footnote{Data Spaces Business Alliance: \url{https://data-spaces-business-alliance.eu/}} was created. DSBA is made up of GAIA-X ISBL\footnote{GAIA-X ISBL: \url{https://gaia-x.eu/}}, Big Data Value Association (BDVA)\footnote{BDVA: \url{https://www.bdva.eu/}}, FIWARE Foundation \footnote{FIWARE Foundation: \url{https://www.fiware.org/}}, and the International Data Spaces Association (IDSA)\footnote{IDSA: \url{https://internationaldataspaces.org/}}. DSBA contributes to the Data Spaces Support Center in coordinating and supporting all European Common Data Spaces in different domains.

\subsection{Related concepts}

There are different concepts related to the OD ecosystem and, while seemingly similar, they carry distinct meanings. Public Data are data that anyone can access, while OD is a subset of Public Data that is offered in a way that is understandable by people and machines~\cite{transparency_by_design_what_is, linked_data_design}. Linked Data are data identified by Uniform Resource Identifiers (URIs), accessible through the Hypertext Transfer Protocol (HTTP), modeled through triples using standards such as RDF and SPARQL, and interconnected with other data by using their respective URIs~\cite{linked_data_the_story}. Linked Open Data are data that fulfill all the principles of Linked Data and Open Data~\cite{linked_data_design}. Any combination can exist, such as Open but non-Linked Data, Linked but non-Open Data, or Linked and Open Data.

OD is accompanied by information that describes the dataset, known as metadata. 
Metadata have a dual purpose. On the one hand, they are used to identify and characterize sources of information, such as data sets, books, or scientific publications. On the other hand, they ease the interoperability between systems through the use of common ontologies. Examples of metadata usage include bibliographic references that enable the identification of scientific publications~\cite{the_journal_coverage}, ontologies to describe the research process~\cite{gangemi2017publishing}, or the use of metadata to register the contribution of artificial intelligence systems in scientific works~\cite{10380268}.
In the same way, metadata within the Open Data ecosystem provides context to the dataset, such as authorship, datetime information, keywords, or spatial information. In addition, it facilitates the search, filtering, and interchange of resources. Instead of directly exchanging the data, metadata is exchanged, containing the endpoint where the original data source can be found. This process, known as harvesting~\cite{lagoze2003making}, enables scalable information exchange while avoiding data inconsistencies. It is scalable because, in most cases, metadata occupies less space than the data and avoids data inconsistencies, since the information is only located in the original data source, avoiding duplication.

Data is the raw material from which information and knowledge can be derived. Data quality is crucial, as making decisions based on poor-quality data can lead to incorrect results. In assessing data quality, two different aspects have to be taken into account. On the one hand, we can evaluate the data based on several dimensions such as accuracy, completeness, timeliness, precision, usability, etc., as proposed in several studies~\cite{MARTIN2023100779, beyond_accuracy}. On the other hand, we can also assess metadata, making use of well-known methodologies such as FAIR. In this paper, when we refer to data quality, we focus mainly on metadata.

The FAIR Principles~\cite{wilkinson2016fair} establish clear and quantifiable criteria for guiding data publications to achieve Findability, Accessibility, Interoperability, and Reusability (FAIR). These principles deliberately avoid delving into contentious issues, such as the specific technology or implementation approach. This high level of abstraction has led to widespread acceptance by numerous research funding organizations and policymakers. The application of FAIR principles in OD contributes to a more efficient, collaborative, and transparent data ecosystem, which benefits both the scientific community and society at large.

\subsection{Open Data Portals}

The emergence of OD led to the deployment of Open Data Portals (ODPs), that is, online repositories with filtering and searching capabilities of OD. According to Opendatasoft in 2024 there are more than 3,700 Open Data Portals around the world~\cite{opendatasoft}. The large number of ODPs makes it difficult to find a specific resource. Harvesting ODPs are portals that act as aggregators of other portals~\cite{lagoze2003making}. They only store the metadata of the original portals, provide resource search capabilities, and point to the originating portal that contains the data~\cite{realizing_the_innovation}. An example of an aggregator portal would be Data.gov\footnote{Data.gov: \url{https://data.gov}}, the US Government’s Open Data Portal that includes more than 250,000 datasets from 131 organizations (December, 2023).    


Data publishers, ODPs, and data consumers are the main stakeholders in the OD ecosystem. However, Immonen et al.~\cite{requirements_of_an_OD} make a finer division and identify 5 different actors: (1) Data providers, which are organizations (public or private) offering data (free or paid); (2) service providers, who offer tools for OD processing; (3) application developers, who create specific applications using OD; (4) end-users of the applications; and (5) infrastructure and tool providers, who offer tools for OD publication and consumption (e.g., marketplaces, application development platforms, cloud service providers, etc.).

The OD lifecycle begins when data providers generate new data (raw data). These data can come from heterogeneous sources, such as web applications, IoT devices, and multimedia files. In this initial phase, data providers determine which data they want to share. Then, service providers harmonize the information, making it accessible and understandable for application developers. Service providers maintain the OD, offering mechanisms for querying and consuming them. Once OD are published, developers can discover, explore, and ultimately exploit them for developing OD-based applications.

\subsection{Enabling technology}
Enabling tools for the development of ODPs include Open Source Data Management Systems (ODMS)~\cite{realizing_the_innovation}. These systems provide capabilities for managing OD, such as publishing, updating, consuming, user and organization management, and metadata definition. When implementing an ODP, there are two main alternatives: using ODMS or building it from scratch~\cite{requirements_of_an_OD}. There is also a hybrid solution that involves starting with an ODMS and customizing it to the specific needs of each ODP. Examples of ODMS include CKAN\footnote{CKAN: \url{http://ckan.org}}, adopted, for example, by Data.gov.es; Socrata\footnote{Socrata: \url{https://dev.socrata.com}}, adopted by the New York City ODP\footnote{NYC ODP: \url{https://opendata.cityofnewyork.us}}; DKAN\footnote{DKAN: \url{https://dkan.readthedocs.io}}, adopted by the Oklahoma ODP\footnote{Oklahoma ODP: \url{https://data.ok.gov}}; or OpenDataSoft\footnote{OpenDataSoft: \url{https://www.opendatasoft.com}}, adopted by the Paris ODP\footnote{Paris ODP: \url{https://opendata.paris.fr}}. Ojo et al.~\cite{realizing_the_innovation} conducted a comparison of different platforms and concluded that CKAN, DKAN, and Semantic MediaWiki\footnote{Semantic MediaWiki: \url{https://www.semantic-mediawiki.org}} are the most comprehensive options, highlighting their open source nature and easy extensibility.

Another enabling technology of OD are the metadata, which are defined as information about the datasets. Metadata ensure the usefulness, discoverability, and usability of datasets. OD can be presented in various formats (JSON, CSV, RDF, etc.) and represented using different vocabularies (the same concept can be modeled in different ways). Searching for datasets within the raw data is not a straightforward task. As a solution, OD includes metadata to facilitate the discovery of datasets. ODMS offer their own metadata schema, but a common practice is to model metadata following the DCAT standard~\cite{dcat_2}. DCAT is an RDF vocabulary developed by W3C that defines classes and properties to model the metadata of data catalogs published on the Web but does not specify the cardinality of relationships or the mandatory properties for each class. DCAT can be extended through profiles by defining additional properties, relationships, and classes, including cardinality, and property constraints. Examples of such extensions are DCAT-AP~\cite{dcat_ap_specification} (used by European data portals) and DCAT-US~\cite{dcat_us} (used by US data portals). Profiles can also extend other profiles, such as DCAT-IT, the profile for Italian data portals that extends DCAT-AP.

As mentioned above, the growth of ODPs has led to the emergence of aggregator portals that harvest smaller portals to facilitate dataset search~\cite{a_systematic_review, van2004resource}. Examples of aggregators are data.gov, that was launched in 2009, and the EU Open Data Portal, which was launched in 2013, both harvesting ODPs in the US and Europe, respectively. There are different techniques for metadata harvesting. One method is the direct download of metadata from files (dump)~\cite{morsey2012dbpedia}. Dump lacks scalability, pagination, or filtering capabilities. Web scraping is another method that involves extracting data directly from web pages~\cite{8782469}. It is a costly and less scalable option, as the aggregator portal has to adapt to each data source and generate the metadata. Another method involves using custom APIs for data or metadata exchange (e.g., the CKAN API). In this case, the aggregator portal must establish connections with each API and be compatible with the technology and formats of the aggregated portals. Finally, the OAI-PMH protocol aims to standardize data collection, working as a client-server architecture over HTTP and XML~\cite{van2004resource, the_oai_pmh}. In this scenario, the aggregator portal acts as the client, and the aggregated portal as the server. Although OAI-PMH does not specify a vocabulary, it is common to use DCAT (or one of its profiles) for metadata exchange. This method improves communication, interoperability, scalability, and offers filtering and pagination capabilities, but requires portals to use the same vocabulary or an adapter to standardize metadata. Table~\ref{table:harvesting_methods} provides a comparative overview of the various harvesting methods.

\begin{table}[!t]
\caption{Comparative of different harvesting methods.}
\label{table:harvesting_methods}
\small
\begin{tabular*}{20pc}{@{}|p{75pt}<{\raggedright}|p{25pt}|p{32pt}|p{23pt}|p{23pt}|@{}}
\hline
& Dump & Web Scraping & API$^{\mathrm{a}}$ & OAI-PMH  \\
\hline
Scalability & & & \checkmark & \checkmark \\
Filtering & \checkmark & & \checkmark & \checkmark \\
Pagination & & & \checkmark & \checkmark \\
DCAT compliant & \checkmark & & \checkmark & \checkmark \\
Standardization & & & & \checkmark \\
Ease for harvested$^{\mathrm{b}}$ & \checkmark & \checkmark & & \\
Interoperability & & & \checkmark & \checkmark \\
\hline
\multicolumn{5}{@{}p{20pc}@{}}{$^{\mathrm{a}}$: It depends on the implementation of the API and the capabilities it offers.} \\
\multicolumn{5}{@{}p{20pc}@{}}{$^{\mathrm{b}}$: Harvested portals do not need any system for been harvested}
\end{tabular*}
\end{table}

Once ODPs are aggregated, the harvesting ODP must offer search and filtering capabilities of datasets using their metadata. The most common alternatives are through the ODP web interface with simple filters based on the title, keywords, source ODP, description, etc. If the metadata are modeled in RDF and stored in a triple store such as Virtuoso\footnote{Virtuoso: \url{https://virtuoso.openlinksw.com}} or AllegroGraph\footnote{AllegroGraph: \url{https://allegrograph.com}}, more advanced searches can be performed using the RDF graph query language SPARQL. For SPARQL-based search to be effective, the metadata must be standardized. DCAT profiles may include controlled vocabularies, which are lists of URIs that a DCAT property can contain, following the principles of Linked Data. For example, the controlled vocabulary for countries in DCAT-AP includes a set of URIs that allow to identify each country.

\section{Role of OD in Europe}
\label{ch:od_europe}

\subsection{European ODPs}

As a consequence of Directives 2003/98/EC~\cite{directive_2003} and 2013/37/EU~\cite{directive_2013}, two initiatives emerged in Europe. In 2012, appeared the EU Open Data Portal for the publication of OD from European Union institutions, bodies, and agencies. In 2015 the European Data Portal for the publication of OD from other existing European portals was launched. In 2021, after the approval of the directive 2019/1024~\cite{directive_2019}, data.europa.eu arises unifying the EU Open Data Portal and the European Data Portal.

Data.europa.eu acts as an aggregator portal. In 2023 it harvested 182 catalogs, including more than 1.6 million of datastest, from 36 countries. 
Data.europa.eu is a repository of metadata modeled through the DCAT-AP specification (the European DCAT profile~\cite{dcat_ap_specification}). It offers different integration mechanisms for data publishers: 

(1) The publisher provides the metadata in DCAT-AP format (Data.europa.eu supports OAI-PMHv2). In this case, data.europa.eu acts as an OAI-PMHv2 client, and the harvested data portal as the OAI-PMHv2 server. Alternatively, data.europa.eu can harvest DCAT-AP metadata through queries in the publishers’ SPARQL endpoints. The publisher also has the possibility to provide the catalog metadata in RDF/XML format directly. 

(2) Direct integration via CKAN. Publishers give data.europa.eu access to their CKAN without the need to be DCAT-AP compliant. Data.europa.eu performs the transformation of metadata in CKAN format to DCAT-AP. 

(3) Integration through an OAI-PMH interface for Infrastructure for Spatial Information in Europe (INSPIRE) and Catalogue Service for the Web (CSW) geospatial catalogs. In data.europa.eu the harvested metadata are transformed into RDF triples. The objective of data.europa.eu is to guarantee the quality of the datasets it harvests. During the harvesting process, it performs a transformation of the metadata to improve its quality. These transformations include automatic translation into the official languages of the European Union. Data consumers do not access data sources directly from data.europa.eu, as the European portal only saves the metadata and not the data. Data.europa.eu helps data consumers find OD sources offering different mechanisms to query them through their metadata. In this way, customers can access catalog metadata through the web interface, the data.europa.eu API, or by using the SPARQL client that searches the RDFs harvested by data.europa.eu. To ease the discovery of datasets it is necessary to define the metadata in a common format. Vocabularies such as DCAT facilitate this task. The DCAT-AP profile adds restrictions to DCAT such as the use of controlled vocabularies, i.e., a set of values that a property can take. For example, the \texttt{theme vocabulary} includes the different themes to which a dataset can be assigned.  

Data.europa.eu also evaluates the quality of the metadata published through the Metadata Quality Assesment (MQA) service. Metadata quality plays a crucial role in ensuring the usefulness, discoverability, and usability of datasets. High-quality metadata provides accurate and comprehensive information about the dataset, facilitating efficient search, discovery, and integration of data across different portals and platforms. The MQA methodology defines a set of indicators to evaluate the harvested catalogs. These indicators encompass various aspects of metadata quality, mainly those based on the FAIR principles. Each indicator is designed to measure specific attributes or characteristics of the metadata, such as the presence of essential information (title, description, keywords), proper licensing and usage rights, temporal and spatial coverage, data format, and access to dataset.
To assess the quality of metadata, the MQA methodology specifies conditions that must be met for each indicator, leveraging the properties provided by the DCAT-AP vocabulary. This ensures that the metadata aligns with a standardized and interoperable representation, enhancing its consistency and compatibility with other datasets and systems.
Furthermore, the MQA methodology assigns weights to each indicator based on its importance in determining the overall quality of the dataset. This allows data providers to prioritize their efforts in improving the aspects of metadata that have a significant impact on its quality and usability.

\subsection{From OD to Data Spaces}


The concept of Data Spaces has emerged as a pivotal framework in the contemporary landscape of data management, offering a structured and dynamic environment for the aggregation, integration, and utilization of diverse datasets to promote the data economy. This section explores the conceptual components of a Data Space and elucidates their intrinsic relationship with European Open Data initiatives, emphasizing the key role of high-quality metadata in fostering seamless integration.

A Data Space can be conceptualized as a virtual ecosystem wherein diverse datasets coexist, facilitated by standardized protocols, interoperability mechanisms, and governance structures. Unlike traditional data repositories, Data Spaces transcend organizational and domain-specific boundaries, providing a collaborative platform where data from various sources can be shared, accessed, and utilized \cite{nagel2021design}.

Within a Data Space, datasets are not isolated entities but interconnected components that contribute to a holistic data environment. The spatial metaphor encapsulated in the term ``Data Space'' implies a virtual realm where data entities exist, interact, and collectively contribute to the general goal of facilitating data-driven insights and applications.

In this regard, the integration of European Open Data within Data Spaces is a strategic endeavor that takes advantage of the principles of openness, accessibility, and interoperability. Open Data, characterized by its unrestricted accessibility and usability, aligns seamlessly with the core characteristics of Data Spaces. Metadata of Open Data within these spaces act as conduits for the flow of information, breaking down silos and promoting collaboration across sectors, industries, and geographical boundaries by guaranteeing interoperability, accessibility, and findability~\cite{reiberg2022data}.

European Open Data initiatives, promoted by the push for transparency and sharing, find an easy integration and adoption within Data Spaces. The interconnected nature of Data Spaces facilitates the cross-fertilization of diverse datasets, enabling stakeholders to take advantage of the collective intelligence embedded in Open Data for informed decision-making and collaborative endeavors.

The foundational components of a Data Space encompass various stakeholders, governance policies, interoperability standards, and crucially, a robust metadata framework~\cite{otto2022designing}. The relationship between these conceptual components and Open Data is symbiotic, as they collectively contribute to the integration and effective utilization of European Open Data:
\begin{itemize}
    \item \textbf{Data Providers}: Entities contributing datasets to the Data Space, including government agencies, research institutions, and private organizations, play a pivotal role in sharing Open Data for collective benefit.
    \item \textbf{Data Consumers}: Stakeholders leveraging Open Data for analysis, research, and decision-making are key players within the Data Space, driving the utilization and value extraction from diverse datasets.
    \item \textbf{Interoperability Standards}: Common protocols and standards within Data Spaces ensure that Open Data from different sources can seamlessly integrate, promoting cross-domain interoperability and collaboration.
    \item \textbf{Data Governance Policies}: Policies governing the ethical and legal use of data within Data Spaces align with the principles of responsible Open Data stewardship, ensuring transparency, accountability, and user trust.
    \item \textbf{Metadata Framework}: High-quality metadata serves as the linchpin, providing structured information that enhances the discoverability, understanding, and usability of Open Data within the Data Space.
\end{itemize}

In the realm of Data Spaces, the metadata framework plays a pivotal role in facilitating the integration and utilization of Open Data. In particular, standards such as NGSI-LD and DCAT (Data Catalog Vocabulary) contribute significantly to this metadata framework, enhancing the comprehensiveness and effectiveness of data management within the dynamic Data Space environment~\cite{nagel2021design}.

Fiware Orion-LD, as a technological enabler, assumes a critical role in the construction and management of metadata within Data Spaces. Positioned as a foundational component, Orion-LD creates a dynamic environment that fosters the coexistence and interaction of diverse datasets. Its adherence to standardized protocols ensures interoperability, allowing for the integration and exchange of Open Data from various sources. Moreover, Orion-LD contributes significantly to the metadata framework, serving as a pivot for the integration and utilization of datasets. It enhances the discoverability, understanding, and usability of Open Data within the Data Space, providing essential insights for stakeholders making informed decisions.


On the other hand, DCAT represents another important implementation within the metadata framework of Data Spaces\cite{guasch2022semantic}. DCAT contributes to the metadata framework by providing a structured and standardized vocabulary that enables a consistent and coherent representation of Open Data. The use of DCAT enhances the discoverability and accessibility of Open Data within the Data Space, promoting a more interoperable data ecosystem.

Together, implementations like Fiware Orion-LD and DCAT synergistically enhance the metadata framework within Data Spaces. Orion-LD's dynamic capabilities and adherence to protocols complement DCAT's standardized vocabulary, creating a robust and interoperable environment for the integration, management, and utilization of Open Data. Their collaborative role ensures that metadata within Data Spaces is not only comprehensive but also aligned with industry standards.

Also, in the European Open Data Ecosystem emerged initiatives such as the EU Open Data Portal and the European Data Portal, in the Data Spaces Ecosystem emerged initiatives to share data in the business ecosystem. The most relevant European initiatives are International Data Spaces (IDS)\footnote{IDS: \url{https://internationaldataspaces.org/}} and Gaia-X\footnote{Gaia-X: \url{https://gaia-x.eu/}}. They define a common framework to establish federated data spaces that preserve the data sovereignty of each participant. Both the Open Data initiatives and the Data Spaces initiatives concur in core concepts and aspects, which foster the extrapolation of the Open Data architectures to the Data Spaces ones.

Among some synergies, we can observe that both ecosystems define similar core roles that interact in the architecture: Data Providers and Data Consumers~\cite{requirements_of_an_OD, otto2022designing}. The IDS, in particular, subdivide these roles to more closely approximate situations occurring in some business data spaces. The IDS also defines a Metadata Broker which acts similar to the metadata aggregator deployed by Open Data initiatives. Likewise, the IDS and the Gaia-X also emphasize interoperability and governance aspects.

The conceptual components of a Data Space and their intricate relationship with European Open Data underscore the significance of fostering a cohesive ecosystem. The collaborative and interconnected nature of Data Spaces, coupled with high-quality metadata practices, paves the way for the effective integration of Open Data, contributing to a more transparent, innovative, and data-driven European landscape.

\subsection{Challenges and opportunities}

In the literature, researchers study the challenges of Data Spaces related to OD barriers that limit their adoption~\cite{wibowo2023systematic, european_strategy_for_data}, highlighting the following regarding data quality, metadata quality, and the OD ecosystem:

\begin{itemize}
    \item Wrong or incomplete data: Incorrect, inaccurate, or incomplete data can lead to incorrect decisions or unreliable results in applications and analyses~\cite{linked_data_in_the_european_data_portal}.
    
    \item Data out of date: Data obsolescence is a common challenge in the OD environment~\cite{9963753}. Outdated data can be irrelevant or even misleading to users.
    
    \item Non-machine-readable datasets: Non-machine-readable formats, such as PDF, make automation and data extraction difficult, limiting their utility in applications that require structured and machine-readable data~\cite{transparency_by_design_what_is}.
    
    \item Low-quality metadata: Incomplete, wrong, and non-update metadata limit the findability, interoperability, accessibility, reusability, and contextuality of datasets~\cite{umbrich2015quality}.
    
    \item Metadata not compliant with the standards: The DCAT standard is widely used to describe datasets, incompatibility of metadata with this standard creates barriers to interoperability and data exchange~\cite{open_data_hopes_and_fears}. The coexistence of multiple DCAT profiles hinders standardization and interoperability, as each profile may have slightly different requirements and structures. 
    
    \item Lack of license information: The lack of clear and adequate license information can pose legal and ethical issues when using OD. The lack of clarity on how data can be used inhibits its adoption by consumers~\cite{overcoming_misattibution}. 
    
    \item Use of commercial formats: Offering data in proprietary or commercial formats, hinders access and usage, as not all users can open or process these formats without additional cost. 
    
    \item Lack of historical data: When OD only provide access to real-time information and lacks historical information limits the ability to conduct retrospective analyses, training of Machine Learning models, etc. 
    
    \item Limited interest from companies: Most OD are offered by public entities, and the lack of interest of private companies to share their data limits the diversity and quantity of available data~\cite{open_data_hopes_and_fears}.
\end{itemize}

Some authors refer to Apparent ODPs as those ODPs that exhibit some of these issues~\cite{criteria_for_odp}. Table \ref{table:barriers_od} summarizes these challenges with real examples (November, 2023).

The European data strategy~\cite{european_strategy_for_data} mentions among the main barriers to the implementation of Data Spaces ensuring data availability, effective data exchange, interoperability and data quality, and the lack of data technology and infrastructure. The synergy between OD and Data Spaces lies not only in the concept, stakeholders, and architecture but also in the barriers they present. Specifically, OD technology acts as an enabler of Data Spaces, but it is necessary to address the challenges mentioned above~\cite{Kirstein2022}.

\begin{table}
\caption{Barriers of OD with examples}
\label{table:barriers_od}
\small
\begin{tabular*}{21pc}{@{}|p{60pt}<{\raggedright}|p{167pt}<{\raggedright}|@{}}
\hline
\multicolumn{2}{@{}|p{20pc}|@{}}{\textbf{Data issues}}\\ \hline
Wrong or incomplete data & The dataset published by the Swiss ODP regarding the availability of electric vehicle charging stations does not include date-time information, making the data unusable$^{\mathrm{a}}$ \\ 
\hline
Outdated data & Real-time air quality dataset from Zaragoza, published by the Spanish ODP, has not been updated since 2021$^{\mathrm{b}}$\\ \hline
Non-machine-readable datasets & The statistical indices about the cause of fires published by the Government Open Data Portal Moldova only offer the OD in PDF format$^{\mathrm{c}}$.  \\ \hline
\multicolumn{2}{@{}|p{20pc}|@{}}{\textbf{Metadata issues}}\\ \hline
Low-quality metadata & The dataset from the Austria ODP related to the evolution of data lakes do not contain a description, geographic information, date-time it was modified, etc.$^{\mathrm{d}}$ \\ \hline
Multiple DCAT profiles & DCAT-US, and DCAT-AP have differences that make them not fully compliant.\\ \hline
Lack of license information & The dataset about the land use in Pays de Lunel published by French ODP does not include information about license$^{\mathrm{e}}$. \\ \hline
\multicolumn{2}{@{}|p{20pc}|@{}}{\textbf{OD ecosystem issues}}\\ \hline
Use of commercial formats & The points of tourist interest dataset published by the Italian ODP only offers an Excel resource $^{\mathrm{f}}$. \\ \hline
Lack of historical data & The parking occupancy dataset of the city of Cologne published by the German ODP only includes real-time information$^{\mathrm{g}}.$ \\ \hline
Limited interest from companies & The Uber Movement Initiative which offered anonymized OD was one of the most important OD sources from private companies. It has been retired in 2023$^{\mathrm{h}}$. \\ \hline
\multicolumn{2}{@{}p{20pc}@{}}{$^{\mathrm{a}}$https://data.europa.eu/data/datasets/ladestationen-oevch~~1}\\
\multicolumn{2}{@{}p{20pc}@{}}{$^{\mathrm{b}}$https://data.europa.eu/data/datasets/https-opendata-aragon-es-datos-catalogo-dataset-oai-zaguan-unizar-es-89320}\\
\multicolumn{2}{@{}p{20pc}@{}}{$^{\mathrm{c}}$https://data.europa.eu/data/datasets/17229-indicii-statistici-despre-cauza-si-locul-izbucnirii-incendiilor-in-republica-moldova}\\
\multicolumn{2}{@{}p{20pc}@{}}{$^{\mathrm{d}}$https://data.europa.eu/data/datasets/unterer-eisbodensee-a-good-example-for-the-future-evolution-of-glacial-lakes-in-austria}\\
\multicolumn{2}{@{}p{20pc}@{}}{$^{\mathrm{e}}$https://data.europa.eu/data/datasets/65307570a1cff0a0856e33b1}\\
\multicolumn{2}{@{}p{20pc}@{}}{$^{\mathrm{f}}$https://data.europa.eu/data/datasets/punti-di-interesse-turistico}\\
\multicolumn{2}{@{}p{20pc}@{}}{$^{\mathrm{g}}$https://data.europa.eu/data/datasets/647ed189-ce31-40db-9b9d-353a7768dadf}\\
\multicolumn{2}{@{}p{20pc}@{}}{$^{\mathrm{h}}$https://www.uber.com/newsroom/introducing-uber-movement-2/}\\
\end{tabular*}
\end{table}

In recent years, countries have made efforts to improve the quality of OD and automate the publishing and harvesting process. By 2023, 82\% of the EU countries indicated that at least 90\% of their datasets include licensing information and offer their datasets under open licenses~\cite{open_data_maturity_in_europe_2023}. This number has increased compared to previous years (in 2016 this percentage was 72\%~\cite{open_data_maturity_in_europe_2016}). Regarding the data published, in 2023 70\% of EU countries report that their datasets include structured data, but only 56\% of their datasets are machine-readable~\cite{open_data_maturity_in_europe_2023}. Although almost half of the resources are still not machine-readable, there has been an improvement compared to previous years. In 2015, only 35\% of the countries published at least half of their datasets that were machine-readable~\cite{open_data_maturity_in_europe_2015}. However, the adoption of LD is still very low (around 22\% in 2023~\cite{open_data_maturity_in_europe_2023}).

All these barriers in the OD need to be addressed in order to effectively implement Data Spaces. The publication of data and the generation of metadata compatible with the DCAT standard limit the involvement of data owners in the Data Spaces ecosystem. New software tools are required to automate the entire process. These tools ensure the quality of the generated metadata, interoperability, accessibility, and scalability, all of them crucial aspects in Data Spaces.

\section{Automatic publication of high-quality OD}
\label{ch:automatic_publication}
Among the barriers mentioned above, the low interest of publishers and the poor quality of the data and metadata stand out. Automating the generation of metadata and the publication of data and metadata speeds up the process, reduces errors, and enables scaling of the OD publication. We have developed software tools that automate the data publication process, including metadata generation and quality assessment.

\subsection{General overview of the solution}

Our proposal is based on three distinct axes. On the one hand, we expand tools for the transformation of the original dataset and metadata generation. ETL (Extract-Transform-Load) systems enable the configuration of a chain of transformations on the original dataset. Our proposal is based on Apache NiFi\footnote{Apache NiFi: \url{https://nifi.apache.org}}, a dataflow system that executes scalable ETL processes through directed graphs. The second axis comprises a data management system. Specifically, we have extended the open-source data management system CKAN to handle both data and metadata. The final axis aims to enable interoperability of the open data manager with other systems and facilitate harvesting while ensuring the quality of metadata. To achieve this, we have developed tools that adapt metadata to different versions and profiles of the DCAT standard compatible with the OAI-PMH catalog exchange protocol, along with tools that allow real-time evaluation of metadata quality. The combination of these three axes enables the automation of the entire process, from data acquisition to its publication and exchange with other Open Data Portals (ODPs). Additionally, our proposal is compliant with DCAT, NGSI-LD the FIWARE software tools, all relevant technologies for the development of Data Spaces~\cite{nagel2021design}.

To explain all the phases of the automatic publication process, we will showcase a real example from a data source provided by the Spanish Meteorological Agency (AEMET). Specifically, this involves the weather station located in Madrid's Retiro Park, which records information such as temperature, precipitation, wind direction, etc. In this use case, the data source is from an external organization, but it can also encompass proprietary data or data from other ODPs.

\begin{itemize}
    \item \textbf{Connection to data sources}. Subscription mechanisms ensure consistency between the original data source and the published data. These mechanisms would notify the publisher system when any changes occur in a dataset. If the original data source does not implement these services, polling with a period significantly shorter than the average update period of the source minimizes inconsistency times. In the case of AEMET, which does not implement these mechanisms but updates the data hourly, a request is made every five minutes to check for changes in the data source. NiFi was used as a tool for data ingestion and polling of the data source.
    
    \item \textbf{Selection of information to be published as OD}. Once raw data are acquired, it is necessary to enable data selection capabilities for the information intended to be offered as OD. NiFi graphs allow to transform origin data, eliminating irrelevant or sensitive information.

    \item \textbf{Enhancement of OD quality}. Many data offered as OD do not adhere to FAIR principles. In the case of Madrid Retiro, data are modeled without following any standard. A NiFi processor has been configured to align the data with the NGSI-LD~\cite{ngsild} standard and the WeatherObserved smart data model. Both the NGSI-LD and Fiware Smart Data models are reference technologies for the development of smart solutions in smart cities~\cite{9963753}.

    \item \textbf{Generation of metadata}: By default, most of the datasets lack metadata, which are necessary for publication in ODPs. In this phase, a NiFi processor has been developed to automatically generate metadata from the AEMET data modeled in the NGSI-LD format.

    \item \textbf{Evaluation of metadata quality}. In this phase, preceding the publication of Open Data, the quality of metadata is assessed. We have developed a tool that evaluates metadata according to criteria defined by data.europa.eu. Data.europa.eu conducts metadata assessment after publication. Pre-evaluation prevents issues in integration with data.europa.eu and ensures high-quality metadata publication since the first aggregation.

    \item \textbf{Enhancement of metadata quality}. Based on the results obtained from the metadata evaluation, detected issues can be rectified. The evaluation and improvement cycle can be iterated multiple times until the desired score is achieved.

    \item \textbf{Publication of OD}. A NiFi processor has been developed to automate the publication of datasets and metadata in CKAN-based ODPs.

    \item \textbf{Interoperability with other ODPs}. A CKAN extension has been implemented to make metadata compatible with different versions of DCAT-AP and the OAI-PMH v2 protocol for future integration into other ODPs such as data.europa.eu.
\end{itemize}

\subsection{Generation of OD and metadata}

As we have mentioned, the initial phases of our proposal involve obtaining raw data, selecting the best-value data, eliminating sensitive information, and improving data modeling. In the case of Madrid Retiro, the transformation has been carried out using the NiFi JOLT processor\footnote{NiFi JOLT processor: \url{https://nifi.apache.org/documentation/nifi-2.0.0-M1/components/org.apache.nifi/nifi-standard-nar/2.0.0-M1/org.apache.nifi.processors.standard.JoltTransformJSON/}}, capable of generating JSON or JSON-LD documents with a specific schema. In this case, an NGSI-LD entity is obtained as a result.

The next step involves the generation of metadata. For this purpose, we have developed the UpdateCKANMetadata processor\footnote{UpdateCKANMetadata processor: \url{https://github.com/ging/fiware-draco/blob/master/docs/processors_catalogue/update_ckan_metadata.md}}. With this processor, we can generate metadata using a template. There are three ways to generate metadata: 1) extracting from the original dataset (e.g., the dataset's title); 2) generating the metadata from the context (e.g., metadata update date); and 3) adding the metadata manually (e.g., license). The processor allows configuring the template once and using it across multiple data sources. In this case, it was used to generate metadata from more than one hundred weather stations managed by AEMET.

\begin{lstlisting}
<?xml version="1.0" encoding="utf-8"?>
<rdf:RDF
  xmlns:vcard="http://www.w3.org/2006/vcard/ns#"
  xmlns:dct="http://purl.org/dc/terms/"
  xmlns:owl="http://www.w3.org/2002/07/owl#"
  xmlns:foaf="http://xmlns.com/foaf/0.1/"
  xmlns:dcat="http://www.w3.org/ns/dcat#"
  xmlns:rdf="http://www.w3.org/1999/02/22-rdf-syntax-ns#"
  xmlns:ns1="http://data.europa.eu/r5r/"
  xmlns:skos="http://www.w3.org/2004/02/skos/core#"
>
  <dcat:Dataset rdf:about="https://portal-yoda.dit.upm.es/dataset/0d7509d6...">
    <dcat:keyword>weather</dcat:keyword>
    <dcat:landingPage>...</dcat:landingPage>
    <dct:spatial>...</dct:spatial>
    <dct:identifier>0d7509d6...</dct:identifier>
    <dct:modified rdf:datatype="http://www.w3.org/2001/XMLSchema#dateTime">2022-10-30T10:05:26.881274</dct:modified>
    <dct:title>madrid_retiro_3195</dct:title>
    <dct:publisher>...</dct:publisher>
    <owl:versionInfo>1.0</owl:versionInfo>
    <dct:temporal>...</dct:temporal>
    <dct:description>Weather of the MADRID RETIRO station...</dct:description>
    <dcat:contactPoint rdf:resource="https://yoda.dit.upm.es/"/>
    <dct:accessRights>...</dct:accessRights>
    <dcat:distribution>...</dcat:distribution>
    <dct:issued rdf:datatype="http://www.w3.org/2001/XMLSchema#dateTime">2022-07-18T20:08:31.590236</dct:issued>
    <dcat:theme>
      <skos:Concept rdf:about="http://publications.europa.eu/resource/authority/data-theme/ENVI"/>
    </dcat:theme>
  </dcat:Dataset>
    ...
</rdf:RDF>

\end{lstlisting}

\subsection{Enhancement of the metadata}

As mentioned in Section~\ref{ch:od_europe}, MQA systems allow data providers to assess the quality of metadata before they are harvested by data.europa.eu, providing them with the opportunity to improve their quality. To implement the MQA functionality, data.europa.eu offers an online tool to validate DCAT-AP files through a set of Shapes Constraint Language (SHACL) specifications. This tool assesses the vocabulary and syntax of metadata within the dataset, ensuring its conformity to defined standards. However, it does not calculate the total score, which hampers data providers' ability to determine the overall quality level of their datasets.

To address this gap, we have developed the mqa-scoring-api tool\footnote{MQA scoring api: \url{https://github.com/YourOpenDAta/mqa-scoring-api}}. This tool verifies whether the requirements specified by the MQA are met and provides the total score, along with a detailed report of the metrics that have not been fulfilled for each indicator within the dataset. By having access to the total score and individual metric evaluations, data providers can gain insight into the specific areas where improvements are needed, enabling them to enhance the quality of their datasets before they are harvested by aggregator portals.

The development of the mqa-scoring-api tool is primarily based on the RDFLib library, a Python package designed for RDF manipulation. RDFLib provides a range of functionalities for working with RDF data, including parsing and serializing different RDF formats, such as RDF/XML. Additionally, RDFLib's Graph object serves as a Python collection of RDF triples (Subject, Predicate, Object), enabling efficient handling and analysis of metadata representations.

To expose the functionality of the mqa-scoring-api tool through an API, we have also developed a Flask server that integrates the MQA tool. 

By focusing on improving metadata quality, data providers can improve the discoverability, trustworthiness, and reusability of their datasets, fostering more effective data sharing, collaboration, and innovation in the open data ecosystem.
We will now show how the MQA Scoring tool works, describing the evolution of the quality of a dataset from AEMET, from its creation to the final version published in data.europa.eu.

In the Madrid Retiro dataset the results of the scoring tool evaluate the dataset with 265 points out of 405 possible. This was mainly because the system which stored the data, did not respond to HTTP HEAD requests. The report generated by the MQA scoring tool states that ``Responded status code of the HTTP HEAD request is not in the 200 or 300 range. No weight assigned''. The MQA tool also reported incompatibilities with DCAT. In this case, SEMIC is in the process of changing SHACL validation shapes of DCAT-AP and therefore did not validate all the metadata correctly. Consequently, incompatibilities with DCAT are a common issue among all catalogs harvested on data.europa.eu.

Taking into account all the issues detected with the MQA Scoring tool, the corresponding fixes were made in the metadata of the AEMET dataset. The scoring tool was then rerun with a much better result (375 out of 405 points). Table \ref{table:madrid_retiro} summarizes the results of both evaluations and the maximum punctuation obtained per dimension. As mentioned above, the incompatibilities with DCAT cannot be solved until the SHACL validation shapes are fixed by SEMIC.


\begin{table}[!t]
\caption{Madrid Retiro dataset metadata quality}
\label{table:madrid_retiro}
\centering
\begin{tabular}{|c|c|c|c|}
\hline
\textbf{Dimension} & \textbf{Punctuation} & \textbf{1$^{\mathrm{st}}$ evaluation} & \textbf{2$^{\mathrm{nd}}$ evaluation}\\
\hline
\multicolumn{4}{|c|}{\textbf{\textit{Findability (100 points)}}} \\ \hline
Keyword usage & 30 & \checkmark & \checkmark \\
Theme usage & 20 & \checkmark & \checkmark \\
Geo info. & 20 & \checkmark & \checkmark \\
\hline
\multicolumn{4}{|c|}{\textbf{\textit{Accessibility (100 points)}}} \\ \hline
Access URL & 50 & X & \checkmark \\
Download URL & 50 & X & \checkmark \\
\hline
\multicolumn{4}{|c|}{\textbf{\textit{Interoperability (110 points)}}} \\ \hline
Format & 25 & \checkmark & \checkmark \\
Media type & 15 & \checkmark & \checkmark \\
Open format & 20 & \checkmark & \checkmark \\
Machine-readable & 20 & \checkmark & \checkmark \\
DCAT compliant & 30 & X & X \\
\hline
\multicolumn{4}{|c|}{\textbf{\textit{Reusability (75 points)}}} \\ \hline
License & 30 & \checkmark & \checkmark \\
Public Access & 15 & \checkmark & \checkmark \\
Contact Point & 20 & \checkmark & \checkmark \\
Publisher info. & 10 & \checkmark & \checkmark \\
\hline
\multicolumn{4}{|c|}{\textbf{\textit{Contextuality (20 points)}}} \\ \hline
Rights & 5 & \checkmark & \checkmark \\
Size & 5 & \checkmark & \checkmark \\
Date info. & 10 & X & \checkmark \\ 
\hline
\textbf{Total} & \textbf{405} & \textbf{265} & \textbf{375} \\
\hline

\end{tabular}
\end{table}

\subsection{Publication of OD}
The final step is the official publication of the OD. We developed the NiFi NGSIToCKAN processor\footnote{NGSIToCKAN processor: \url{https://github.com/ging/fiware-draco/blob/master/docs/processors_catalogue/ngsi_ckan_sink.md}} for carrying out the publication. This processor interacts with the CKAN API, creating all necessary data structures (organizations, packages, and resources), and publishes the metadata following the CKAN format. The processor can be configured to publish all the data or indicate the URI where to find them. In the Madrid Retiro example, the data are not directly published in the ODP; instead, it remains within the Orion-LD Context Broker\footnote{Orion Context Broker: \url{https://github.com/FIWARE/context.Orion-LD}}, a server that is compatible with the NGSI-LD standard with notification capabilities, and an enabling technology for the development of the Data Spaces and OD~\cite{Ahle2022, collaboration_of_dts}

Lastly, we developed the CKAN ckanext-dcatapedp module\footnote{ckanext-dcatapedp extension: \url{https://github.com/YourOpenDAta/ckanext-dcatapedp}} that extends the ckanext-dcat\footnote{ckanext-dcat extension: \url{https://github.com/ckan/ckanext-dcat}} and the ckanext-oaipmh\footnote{ckanext-oaipmh extension: \url{https://extensions.ckan.org/extension/oaipmh/}}  modules to make the metadata compatible with DCAT-AP versions 2.0.1 and 2.1.0, and to be served using the OAI-PMH v2 protocol. The final Madrid Retiro dataset used as an example was harvested by data.europe.eu\footnote{Madrid Retiro dataset: \url{https://data.europa.eu/data/datasets/0d7509d6-82ca-4dc3-a61a-95e45dab7a1e}, \url{https://zenodo.org/doi/10.5281/zenodo.10636027}} getting a final score of 375 points.

\begin{figure*}[!t]
\centerline{\includegraphics[width=0.7\textwidth]{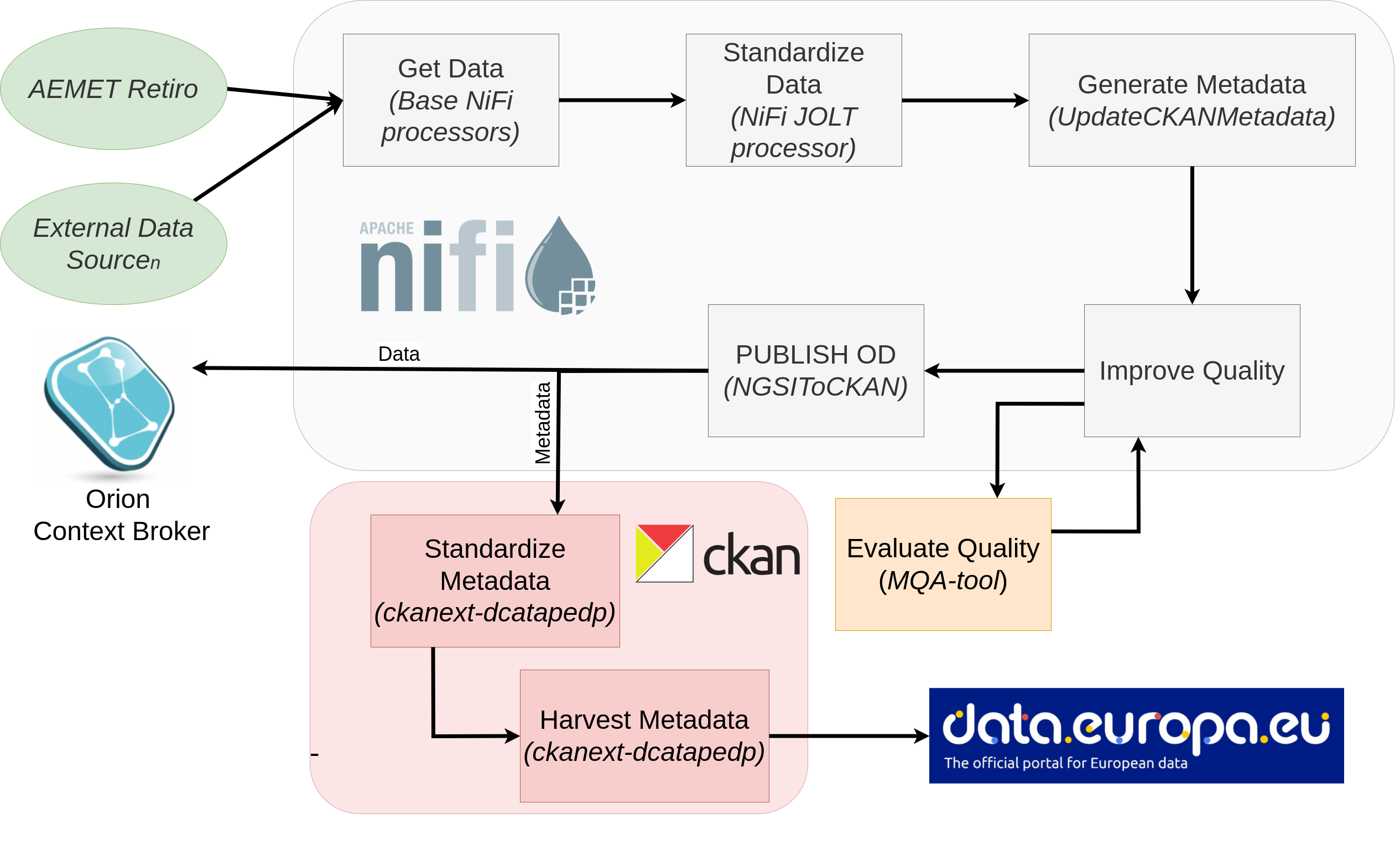}}
\caption{Phases of automatic publication of high-quality OD}
\label{fig:process}
\end{figure*}

Figure~\ref{fig:process} summarizes all the phases and tools required to automate the publication of OD, guaranteeing the quality of the data, metadata, and their interoperability.

\section{Validation and Results}
\label{ch:validation}
We have validated our proposal in the scope of the European project YODA (Your Open DAta\footnote{YODA project: \url{ https://yoda.dit.upm.es/}}). The project was funded by CEF Telecom with the objective of promoting OD and facilitating its use. In the scope of YODA, we implemented the YODA ODP\footnote{YODA Portal: \url{https://portal-yoda.dit.upm.es/en/}} where all the OD are published. The information published in YODA comes from different services of different nature including information on mobility, public services, weather, and pollution. During the research conducted in YODA, over two hundred datasets have been published and harvested by data.europa.eu yielding high-quality results.

\subsection{Scaling-up the solution}

The published datasets are sourced from three different organizations. 1) AEMET which provides meteorological data. AEMET registers meteorological information from a vast number of stations spread throughout the country. AEMET provides information on weather observation, climatology, radiation, contamination, satellite information, weather forecasts, etc. It offers the possibility to access these data through the AEMET OpenData API. However, their metadata are not DCAT-AP compliant. 2) SmartSantander, which provides information on smart mobility and public services. The data provided by the SmartSantander platform come from the LoRaWAN Smart Parking service. This service has been deployed in two locations in the city of Santander, and is based on the use of parking sensors buried under the asphalt to monitor the occupancy status of parking spots on a 1-to-1 basis (i.e., one parking sensor per parking spot). 3) Smart Campus CEI Moncloa which provides equipment density and population data from the university area of Madrid. The Smart Campus CEI Moncloa is a joint initiative of Complutense and the Universidad Politécnica de Madrid with the aim of sustainably transforming the Moncloa Campus into an international benchmark in research, training, and innovation. We obtain anonymized data provided by a set of WIFI sensors deployed on campus to track student activity and extract statistics. As the data contains sensitive information, they are anonymized before collection.



To facilitate the publication of the aforementioned datasources three configurations have been used, one from each organization. With the AEMET configuration, we integrated 122 datasets, with SmartSantander 60, and with CEI Moncloa 20. Through this research, we validated the scalability of our approach, since a single initial configuration enables the generation of metadata and publication of large volumes of OD. Leveraging Nifi's Big Data processing capabilities, the solution can scale and concurrently ingest numerous datasets. 

\subsection{Results in Data.europa.eu}

\begin{figure}[!t]
\centerline{\includegraphics[width=0.5\textwidth]{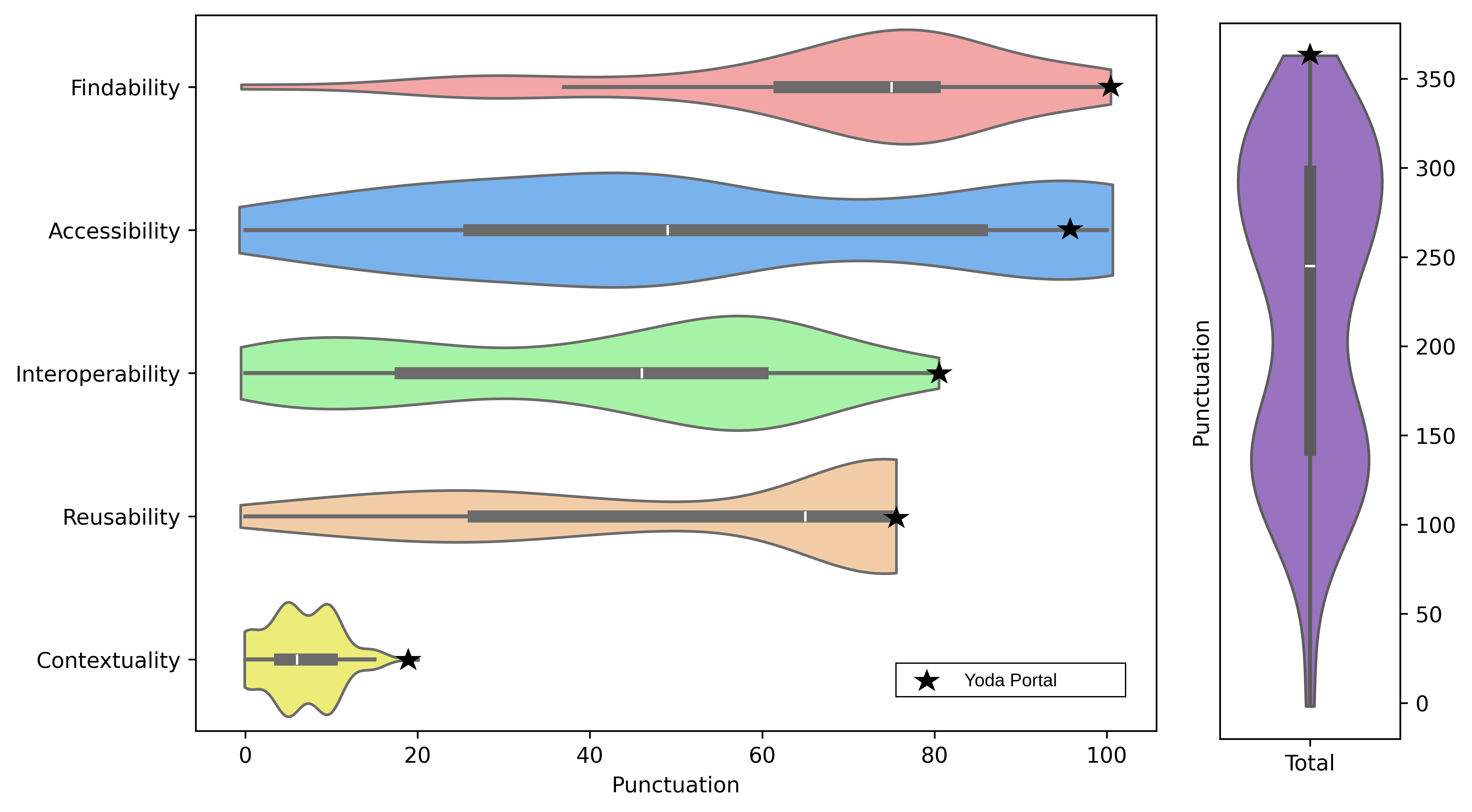}}
\caption{Quality of Open Data Portals harvested by data.europa.eu (November, 2023)}
\label{fig:punctuation}
\end{figure}

All datasets published in the YODA portal have been harvested from data.europa.eu using the OAI-PMH v2 and DCAT-AP v2.1.0 profile implemented with the ckanext-dcatapedp CKAN extension. Figure~\ref{fig:punctuation} presents a comparison of all data catalogs harvested by data.europa.eu\footnote{MQA results data.europa.eu in November 2023: \url{https://zenodo.org/doi/10.5281/zenodo.10636747}}. YODA has a punctuation of 100.0 in terms of findability compared to the rest of the harvested catalogs, which have an average score of 69.3 points (median = 75.0, SD = 22.0). Concerning different dimensions, YODA received a score of 96.0 points in accessibility compared to the average of 53.7 points for the rest of the catalogs (median = 49.0, SD = 32.6); 80.0 points for interoperability compared to 40.2 points (median = 46.0, SD = 24.2); 75.0 for reusability compared to 51.3 points (median = 65.0, SD = 26.6); and 20.0 for contextuality compared to 6.6 (median = 6.0, SD = 4.2) for the rest of the portals.

\section{Conclusions and future work}
\label{ch:conclusions}
In this article, we have analyzed the barriers of OD that limit the adoption of Data Spaces. We presented and proposed a set of tools and procedures to address them. Our proposal includes a set of open source software components for 1) transforming the original dataset removing non-relevant information 2) generating high-quality metadata compliant with DCAT; 3) assessing the quality of metadata according to the specification of the Open Data Portals; and 3) publishing the OD in the corresponding portal. For the sake of understanding, we have illustrated the automatic publication process with a real example from a data source provided by the AEMET. 

We validated our proposal in the scope of the CEF Telecom-funded project Your Open DAta. We have applied our software tools and procedure to automate the publication of over 200 datasets from AEMET, SmartSantander, and Smart Campus CEI Moncloa in the European portal with excellent results, according to the official metrics of data.europa.eu.  

The next step in our research is to focus the proposed solution on providing full integration with the Data Spaces of different verticals. One of the main concerns of this integration is meeting sufficient quality standards according to FAIR principles to ensure the most efficient discovery of data and interoperability with data providers and consumers. Thanks to the proposals presented in this work, we are closer to achieving this objective but there is still an open future work to start testing this approach in real Data Spaces with heterogeneous actors involved. 

\section*{Acknowledgements}

This research is supported by the Eunomia project carried out within the framework of the Recovery, Transformation and Resilience Plan funds, financed by the European Union (Next Generation); and cofinanced by the Connection Europe Facility of the European Union under 2019-ES-IA-0121 of the European Commission: Your Open DAta.

\bibliographystyle{IEEEtran}
\bibliography{ref}



\section{Biography Section}
 



\begin{IEEEbiographynophoto}{Javier Conde} is currently an Assistant Professor at UPM. His research interests include Big Data, Digital Twins, Linked Open Data, Artificial Intelligence, and Data Spaces.
\end{IEEEbiographynophoto}

\begin{IEEEbiographynophoto}{Alejandro Pozo} is currently an Assistant Professor in the Department of Telematics Engineering at UPM. His research interests include Identity Management, Security, Internet of Things, and Multivideoconverencing Systems.
\end{IEEEbiographynophoto}

\begin{IEEEbiographynophoto}{Andres Munoz-Arcentales} is currently an Assistant Professor in the Department of Telematics Engineering at UPM. His main research interests are Data Spaces, Microservices, Data Fusion, Machine Learning, Edge Computing and Big Data.
\end{IEEEbiographynophoto}

\begin{IEEEbiographynophoto}{Johnny Choque} is currently working as a researcher in the Universidad de Cantabria where he received his Ph.D. in 2014. He is active on the IoT-enabled Smart Cities, Low Power Wireless Technologies, Open Data, and Blockchain.
\end{IEEEbiographynophoto}

\begin{IEEEbiographynophoto}{\'Alvaro Alonso} is currently an Associate Professor with the Department of Telematics Engineering at UPM. His main research interests are Public Open Data, Security Management in Smart Context environments, and Multivideoconferencing Systems.
\end{IEEEbiographynophoto}

\vfill

\end{document}